# Robust Peak Detection for Holter ECGs by Self-Organized Operational Neural Networks

Moncef Gabbouj, *Fellow, IEEE*, Serkan Kiranyaz, *Senior Member, IEEE*, Junaid Malik, Muhammad Uzair Zahid, Turker Ince, Muhammad E. H. Chowdhury, *Senior Member, IEEE*, Amith Khandakar, *Senior Member, IEEE*, and Anas Tahir

***Abstract*— Although numerous R-peak detectors have been proposed in the literature, their robustness and performance levels may significantly deteriorate in low-quality and noisy signals acquired from mobile electrocardiogram (ECG) sensors, such as Holter monitors. Recently, this issue has been addressed by deep 1-D convolutional neural networks (CNNs) that have achieved *state-of-the-art* performance levels in Holter monitors; however, they pose a high complexity level that requires special parallelized hardware setup for real-time processing. On the other hand, their performance deteriorates when a compact network configuration is used instead. This is an expected outcome as recent studies have demonstrated that the learning performance of CNNs is limited due to their strictly homogenous configuration with the sole linear neuron model. This has been addressed by operational neural networks (ONNs) with their heterogenous network configuration encapsulating neurons with various nonlinear operators. In this study, to further boost the peak detection performance along with an elegant computational efficiency, we propose 1-D Self-Organized ONNs (Self-ONNs) with generative neurons. The most crucial advantage of 1-D Self-ONNs over the ONNs is their self-organization capability that voids the need to search for the best operator set per neuron since each generative neuron has the ability to create the optimal operator during training. The experimental results over the China Physiological Signal Challenge-2020 (CPSC) dataset with more than one million ECG beats show that the proposed 1-D Self-ONNs can significantly surpass the *state-of-the-art* deep CNN with less computational complexity. Results demonstrate that the proposed solution achieves a 99.10% F1-score, 99.79% sensitivity, and 98.42% positive predictivity in the CPSC dataset, which is the best R-peak detection performance ever achieved.**

Manuscript received January 6, 2021; revised November 10, 2021; accepted March 4, 2022. The work was supported in part by the Academy of Finland through the Project AWcHA under Grant 334566. *(Corresponding author: Moncef Gabbouj.)*

Moncef Gabbouj, Muhammad Uzair Zahid, and Junaid Malik are with the Department of Computing Sciences, Tampere University, 33100 Tampere, Finland (e-mail: moncef.gabbouj@tuni.fi; muhammaduzair.zahid@tuni.fi; junaid.malik@tuni.fi).

Serkan Kiranyaz, Muhammad E. H. Chowdhury, Amith Khandakar, and Anas Tahir are with the Department of Electrical Engineering, College of Engineering, Qatar University, Doha 2713, Qatar (e-mail: mkiranyaz@qu.edu.qa; mchowdhury@qu.edu.qa; amitk@qu.edu.qa; a.tahir@qu.edu.qa).

Turker Ince is with the Department of Electrical and Electronics Engineering, İzmir University of Economics, 35330 İzmir, Turkey (e-mail: turker.ince@izmirekonomi.edu.tr).

## I. Introduction

An ELECTROCARDIOGRAM (ECG) acquires the heartbeat sequence in time displaying the electrical depolarization-repolarization patterns of the heart. ECG signal forms itself in the QRS complexes and ventricular beats and it bears essential information about the status of the heart. Among many other tools, ECG is still the most significant noninvasive tool for cardiac monitoring and clinical diagnosis. R-peak detection is the primary operation that usually precedes any kind of ECG analysis, such as ECG beat classification and cardiac arrhythmia detection [1]–[21]. Conventional Holter monitors and the recent introduction of low-cost and low-power mobile ECG sensors present a significant motive and challenge for robust and real-time detection of the R-peak locations. Especially, robustness is a key issue since it was reported in a recent study [22] that the R-peak detection performance can severely deteriorate when the ECG acquisition is poor and corrupted by a high level of noise.

Particularly for clinical ECG recordings, numerous R-peak detection algorithms have been proposed in the literature. One of the first and the most widely used algorithm was proposed by Pan and Tompkins [23], which has served as the benchmark method for more than three decades. Afterward, several other popular methods based on signal processing have emerged, such as wavelet transform [24], [25], Hilbert transforms [26], and ensemble empirical mode decomposition [27]. Some hybrid methods that consist of traditional signal processing and machine learning have followed, e.g., R-peak detection by radial basis functions (RBFs) [28] and hidden Markov models (HMMs) [29]. The common approach in those classical detectors was to perform R-peak enhancement first by using signal processing techniques, such as filter banks and spectral analysis, and then applying the peak detection. A crucial advantage of such methods is that they are very fast. However, they are all designed for clinical ECG recordings with a clean, almost noise-free signal. All of them were evaluated on the benchmark Massachusetts Institute of Technology-Beth Israel Hospital (MIT-BIH) arrhythmia dataset [35] or other similar datasets with high-quality clinical ECG recordings. Their performance level significantly deteriorates when the ECG signal quality is poor [22], and thus, they are not suitable for low-power mobile ECG sensors. In fact, a proper evaluation constitutes a major problem in general because even though there are few public datasets [30], [31] containing some noisy ECG signals with ground-truth R-peak locations, they are limited in size and duration. This problem has become the main bottleneck even for recent and modern peak detectors based on deep learning paradigms, including long short-term memory (LSTM) networks [32], and convolutional neural networks (CNNs) [33], [34]. Both approaches in [32] and [33] aimed to improve the robustness but only against the artificial (additive) noise. For this purpose, they have only induced certain additive noise, such as baseline wander and motion artifacts, to ECG records in the MIT-BIH dataset and reported a reasonable level of robustness. However, such artificial noise addition does not represent the actual degradations that occur in a Holter monitor where besides the severe and varying noise levels and occasional glitches, the baseline dc level varies drastically and frequently along with the dynamic range of QRS complexes, as shown in Fig. 1. Another issue in such evaluation over the MIT-BIH dataset is that this benchmark dataset does not contain a sufficient number of beats and variations where a deep network can truly be tested. This brings the danger of a certain bias and overfitting the results.

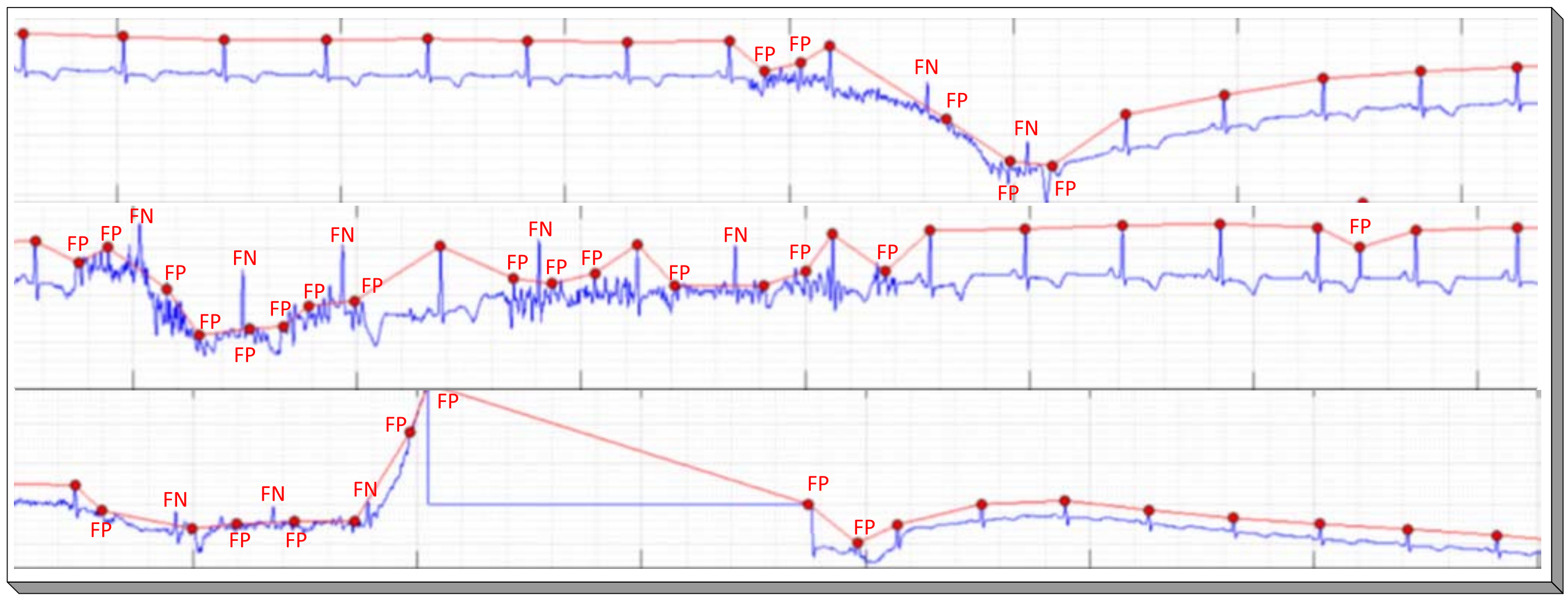

Fig. 1. Typical Holter ECG segments from the record of patient 6 in the CPSC dataset. Red circles represent the R-peaks detected by the Pan–Tompkins method [23] where several FPs and FNs are visible.

Recently, the study in [34] has proposed a novel method based on deep 1-D CNNs that has actually been applied over a real Holter ECG repository, the China Physiological Signal Challenge (2020) dataset (CPSC) with more than 1 million ECG beats. This method has been tested over the entire dataset and achieved the *state-of-the-art* performance level for R-peak detection with a significant margin. However, it still suffers two major drawbacks: 1) 12-layer deep model poses a significant computational complexity, and 2) both false-positives (FPs) and false-negatives (FNs) are still too high, especially on the arrhythmia beats. In this study, we address both issues with a novel network model.

According to the recent studies [42]–[50], the main drawback of the conventional multilayer perceptrons (MLPs) and their derivatives, CNNs, is that they both depend on the ancient neuron model (McCulloch–Pitts) [41]. It is neuroscientific fact that the mammalian neural systems are highly heterogeneous and consist of diverse (nonlinear) neuron types with specialized electrophysiological and biochemical properties [42], [43]. This linear neuron model is only a crude and simplified model of its biological counterpart. As a consequence, MLPs and CNNs with an entirely homogenous network configuration with linear neurons are capable of learning for relatively simple and linearly separable problems; however, they entirely fail to do so whenever the solution space of the problem is highly nonlinear and complex [42]–[50]. To address this major deficiency, generalized operational perceptrons (GOPs) [42]–[49] and later on operational neural networks (ONNs) [50] have been proposed. Both models are heterogeneous with any nonlinear neuron model, and this gives them an elegant diversity level to learn highly complex and multimodal functions or spaces with minimal network complexity and training data. The operational neurons in GOPs and ONNs mimic the biological neurons with, *nodal* (corresponding to the synaptic connections) and *pool* (corresponding to the integration in the soma) operators. An "operator set" is the combination of *nodal*, *pool*, and *activation* operators, and the operator set library is formed in advance to store all potential operator sets. However, ONNs can only achieve a limited heterogeneity level since one operator set has to be assigned to all neurons of each hidden layer. Furthermore, they are bound to the limited number of operators in the operator set library. The latter can be a serious bottleneck on the learning performance when the right operator set needed for the learning problem at hand is missing in the library. Finally, ONNs pose a high computational complexity especially during the training because the right operator sets have to be searched in advance with several backpropagation (BP) runs.

In this study, in order to address the aforementioned issues and drawbacks, we propose 1-D self-organized ONNs (Self-ONNs) with the *generative* neuron model for robust R-peak detection for the Holter ECG in real-time. The generative neuron model enables the self-organization capability for Self-ONNs [51]–[55] where the nodal operators are iteratively generated during the BP training to maximize the learning performance. Obviously, the ability to create *any* nonlinear nodal operator enables superior operational diversity and flexibility. Therefore, Self-ONNs neither need an operator set library in advance, nor require any prior search process to find the optimal nodal operator. The 2-D Self-ONNs have been proposed in recent studies [50] and [54] have shown that even with a few neurons they can achieve a superior learning performance for various image processing and regression tasks while the performance gap between ONNs and CNNs widens further. Hence, in this study, our primary goal is to achieve a superior R-peak detection performance compared with the deep 1-D CNNs [34] while reducing the network complexity and depth significantly for a real-time application. Besides [34], we shall also perform comparative evaluations against the earlier state-of-the-art methods [23], [56]–[59] to accomplish an overall validation. As a summary, we can enlist the novel and significant contributions of this article as follows.

1) This is the first study where 1-D Self-ONNs[1] have even been proposed for ECG peak detection evaluated over the largest ECG benchmark dataset the CPSC (2020) dataset with more than one million ECG beats.
2) Thanks to the heterogeneous network model with generative neurons, 1-D Self-ONNs exhibit a superior learning capability that achieved the *state-of-the-art* peak detection performance even though it has half the network depth and more than four times fewer neurons.
3) The most important contribution of the proposed approach is the crucial reduction achieved on the missed (false-negatives) for arrhythmia beats compared with [34] and other major peak detection methods.
4) Along with superior R-peak detection performance compared with the deep 1-D CNNs [34], the proposed 1-D Self-ONNs enable a significant reduction in the depth and complexity of the deep CNN model proposed in [34].
5) Finally, this is the first article where the raw-vectorized BP formulations are presented for 1-D Self-ONNs along with the corresponding computational complexity analysis.

The rest of the article is organized as follows. Section II presents 1-D Self-ONNs with generative neurons and formulates the forward-propagation (FP) and BP training. Section III outlines the methodology followed in the paper. Section IV describes the ECG datasets and presents the experimental setup used for testing and evaluation. The experimental results and comparative evaluations using standard performance metrics against several state-of-the-art techniques are provided, followed by a detailed complexity analysis, in the same section. Finally, Section V concludes the paper and suggests topics for future research.

[1]The optimized PyTorch implementation of the peak detector based on 1-D Self-ONNs and the benchmark ECG CPSC dataset are publicly shared in https://github.com/MUzairZahid/R-Peak-Detection-1D-SelfONN. The optimized PyTorch implementation of Self-ONNs is freely shared in http://selfonn.net/

## II. 1-D Self-Organized Operational Neural Networks

In this section, we will proceed by revisiting how ONNs generalize the 1-D convolution operation. Then, the mathematical model of the proposed generative neuron-based 1-D Self-ONN will be presented. To conclude, a simplification of the generative neuron will be discussed which can significantly reduce the computational cost by enabling the use of fast vectorized operations.

ONNs are derived from the GOPs in the same way CNNs are derived from MLPs with two restrictions: limited connectivity and weight sharing. GOPs have been proposed in [42] and [45] to replace the basic (linear) neuron model from the 1950s (McCulloch–Pitts) [41] aiming to address the well-known limitations and drawbacks of MLPs. Recently, GOPs have outperformed not only MLPs but even the latest variants of extreme learning machines (ELMs) [46]–[49]. Derived directly from GOPs, ONNs [50] are heterogeneous networks encapsulating neurons with linear and nonlinear operators, hence, carrying a closer link to biological systems. In brief, ONNs extend the sole usage of linear convolutions in the convolutional neurons by the *nodal* and *pool* operators.

Let us consider the case of the $k$th neuron in the $l$th layer of a 1-D CNN. For the sake of brevity, we assume the same convolution operation with unit stride and the required amount of zero padding. The output of this neuron can be formulated as follows:

$$x_k^l = b_k^l + \sum_{i=0}^{N_{l-1}} x_{ik}^l \tag{1}$$

where $b_k^l$ is the bias associated with this neuron and $x_{ik}^l$ is defined as

$$x_{ik}^l = \text{Conv1}D\left(w_{ik}, y_i^{l-1}\right). \tag{2}$$

Here, $w_{ik} \in \mathbb{R}^K$ is the kernel connecting the $i$th neuron of $(l-1)$th layer to the $k$th neuron of the $l$th layer, while $x_{ik}^l \in \mathbb{R}^M$ is the input map, and $y_i^{l-1} \in \mathbb{R}^M$ are the $l$th and $(l-1)$th layers' $k$th and $i$th neurons' outputs, respectively. By definition, the convolution operation of (2) can be expressed as

$$x_{ik}^l(m) = \sum_{r=0}^{K-1} w_{ik}^l(r) y_i^{l-1}(m+r). \tag{3}$$

The core idea behind an operational neuron is a generalization of the earlier as follows:

$$\overline{x_{ik}^l}(m) = P_k^l\left(\psi_k^l\left(w_{ik}^l(r), y_i^{l-1}(m+r)\right)\right)_{r=0}^{K-1} \tag{4}$$

where $\psi_l^k(\cdot): R^{M\times K} \rightarrow R^K$ and $P_k^l(\cdot): R^K \rightarrow R^1$ are termed as *nodal* and *pool* functions, respectively, and assigned to the $k$th neuron of $l$th layer. In a heterogenous ONN configuration, every neuron has uniquely assigned $\psi$ and $P$ operators. Owing to this, an ONN network enjoys the flexibility of incorporating any nonlinear transformation, which is suitable for the given learning problem. However, hand-crafting a suitable library of possible operators and searching for an optimal one for each neuron in a network introduces a significant overhead, which rises exponentially with increasing network complexity.

Moreover, it is also possible that the ideal operator for the given learning problem cannot be expressed in terms of well-known functions. To resolve this key limitation, a composite nodal function is required that is iteratively created and tuned during BP. A straightforward choice for accomplishing this would be to use a weighted combination of all operators in the operator set library and learn the weights during training. However, such a formulation would be susceptible to instability issues because of the different dynamic ranges of individual functions. In addition, it would still rely on the manual selection of suitable functions to populate the operator set library. Therefore, to formulate a nodal transformation that does not require any preselection and manual assignment of operators, we make use of the Taylor-series-based function approximation.

The Taylor series expansion of an infinitely differentiable function $f(x)$ near a point $x = a$ is given as

$$f(x) = \sum_{n=0}^{\infty} \frac{f^{(n)}(a)}{n!}(x-a)^n. \quad (5)$$

The $Q$th order truncated approximation of (5), formally known as the Taylor polynomial, takes the form of the following finite summation:

$$f(x)^{(Q)} = \sum_{n=0}^{Q} \frac{f^{(n)}(a)}{n!}(x-a)^n. \quad (6)$$

The above-mentioned formulation enables the approximation of any function $f(x)$ sufficiently well in the close vicinity of $a$. If the coefficients $(f^{(n)}(a)/n!)$ are tuned and the inputs are bounded, the formulation of (6) can be used to *generate* any transformation. This is the key idea behind the generative neurons which form Self-ONNs. Specifically, in terms of the notation used in (4), the nodal transformation of a generative neuron would take the following general form:

$$\widetilde{\psi_k^l}\left(w_{ik}^{l(Q)}(r), y_i^{l-1}(m+r)\right) = \sum_{q=1}^{Q} w_{ik}^{l(Q)}(r,q)\left(y_i^{l-1}(m+r)\right)^q. \quad (7)$$

In (7), $Q$ is a hyperparameter that controls the degree of the Taylor series approximation, and $w_{ik}^{l(Q)}$ is a learnable kernel of the network. A key difference in (7) as compared with the convolutional (3) and operational (4) model is that $\widetilde{\psi_k^l}$ is not fixed, rather a distinct operator over each individual output, $y_i^{l-1}$, and thus, requires $Q$ times more parameters. Therefore, the $K \times 1$ kernel vector $w_{ik}^l$ has been replaced by a $K \times Q$ matrix $w_{ik}^{l(Q)} \in R^{K \times Q}$ which is formed by replacing each element $w_{ik}^l(r)$ with a $Q$-dimensional vector $w_{ik}^{l(Q)}(r) = [w_{ik}^{l(Q)}(r,0), w_{ik}^{l(Q)}(r,1), \ldots, w_{ik}^{l(Q)}(Q-1)]$. The input map of the generative neuron, $\widetilde{x}_{ik}^l$ can now be expressed as

$$\widetilde{x_{ik}^l}(m) = P_k^l\left(\sum_{q=1}^{Q} w_{ik}^{l(Q)}(r,q)\left(y_i^{l-1}(m+r)\right)^q\right)_{r=0}^{K-1}. \quad (8)$$

During training, as $w_{ik}^{l(Q)}$ is iteratively tuned by the BP, *customized* nodal transformation functions will be generated as a result of (8), which would be uniquely tailored for $ik$th connection. This enables enhanced flexibility which provides three key benefits. First, the need for manually defining a list of suitable nodal operators and searching for the optimal operator for each neuron connection is naturally alleviated. Second, the heterogeneity is not limited to each neuron connection $i \rightarrow k$ but down to each kernel element as $\widetilde{\psi}_l^k(w_{ik}^{l(Q)}(r), y_i^{l-1}(m+r))$ will be unique $\forall r \in [0, 1, \ldots, K-1]$. As show in Fig. 2, such diversity is not achievable even with the flexible operational neuron model of ONNs. Third, in generative neurons, the heterogeneity is driven only by the values of the weights $w_{\mathrm{ik}}^{l(Q)}$ and the core operations (multiplication and summation) are the same for all neurons in a layer, as shown in (8). Owing to this, unlike ONNs, the generative neurons inside a Self-ONN layer can be parallelized much more efficiently, which leads to a considerable reduction in computational complexity and time. Moreover, a special case of (8) can also be expressed in terms of the widely applicable convolutional model.

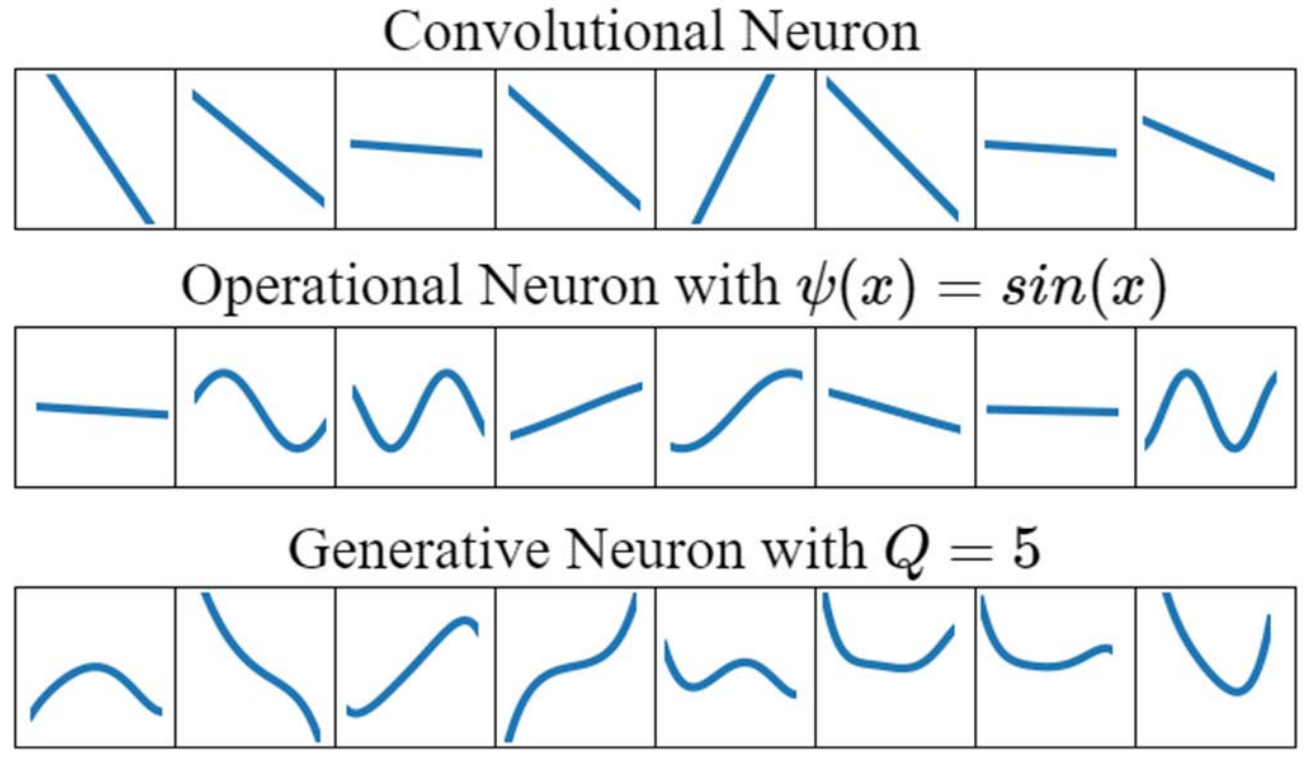

Fig. 2. Visual comparison of different nodal transformation profiles entailed by the kernel of a convolutional, operational, and the proposed generative neuron of order $Q$. The generative neuron model enables enhanced nonlinearity and heterogeneity within the kernels.

### A. Representation in Terms of Convolution

If the pooling operator $P_k^l$ is fixed to summation operator, $\widetilde{x}_{ik}^l$ is then defined as

$$\widetilde{x_{ik}^l}(m) = \sum_{r=0}^{K-1}\sum_{q=1}^{Q} w_{ik}^{l(Q)}(r,q)\left(y_i^{l-1}(m+r)\right)^q. \quad (9)$$

Exploiting the commutativity of the summation operations in (9), we can alternatively write

$$\widetilde{x_{ik}^l}(m) = \sum_{q=1}^{Q}\sum_{r=0}^{K-1} w_{ik}^{l(Q)}(r,q-1)y_i^{l-1}(m+r)^q. \quad (10)$$

Using (1) and (2), the formula in (10) can be further simplified as follows:

$$\widetilde{x_{ik}^l} = \sum_{q=1}^{Q} \mathrm{Conv1}D\left(w_{ik}^{l(Q)}, \left(y_i^{l-1}\right)^q\right). \quad (11)$$

Hence, the formulation can be accomplished by applying Q 1-D convolution operations. If $Q$ is set to 1, (11) entails the convolutional formulation of (3). Therefore, as CNN is a subset of ONN corresponding to a specific operator set, it is also a special case of Self-ONN with $Q = 1$ for all neurons.

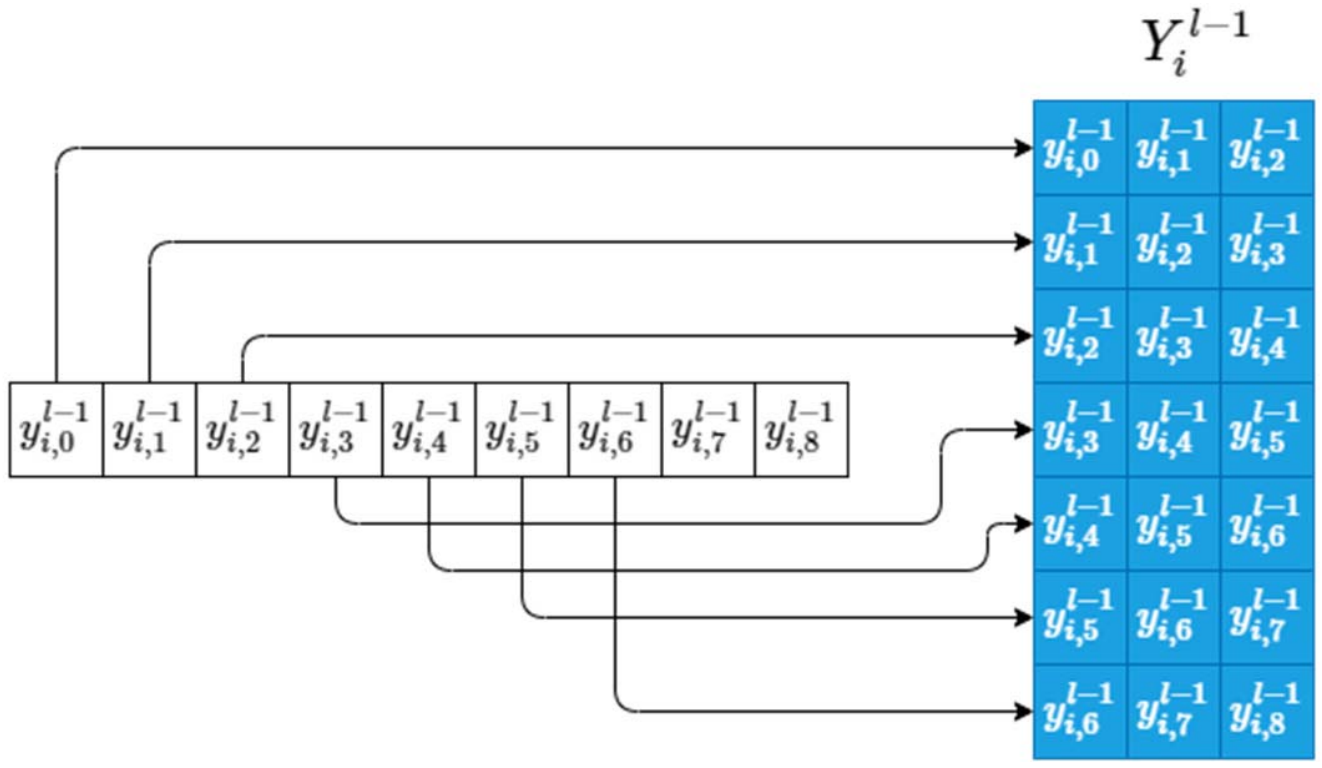

Fig. 3. Reshuffling operation used to convert $\mathbf{y_i^{l-1}}$ to $\mathbf{Y_i^{l-1}}$.

### B. Vectorized Notation

Expressing explicit loops in terms of matrix and vector manipulations is a key idea behind vectorization, which is a major driving factor behind fast implementations of modern-day neural network implementations. In this section, we first introduce how the vectorized notation can be used to express the 1-D convolution operation inside a neuron. Afterward, the same key principles will be exploited to express the generative neuron formulation of (9) as a single matrix-vector product.

First, an alternate formulation of the operation of (3) is now presented. We introduce a transformation $\delta(\cdot, K)$, which concatenates $y_i^{l-1}$ such that values inside each $K$-dimensional kernel as rows to form a matrix $Y_i^{l-1} \in R^{M\times K}$. The process is visually depicted in Fig. 3 for $K = 3$, and mathematically expressed in (12)

$$
\begin{aligned}
&Y_i^{l-1}\\
&= \delta\left(y_i^{l-1}, K\right)\\
&= \begin{bmatrix}
y_i^{l-1}(0) & y_i^{l-1}(1) & \dots & y_i^{l-1}(K-1)\\
\vdots & \vdots & \dots & \vdots\\
y_i^{l-1}(m) & y_i^{l-1}(m+1) & \dots & y_i^{l-1}(m+K-1)\\
\vdots & \vdots & \dots & \vdots\\
y_i^{l-1}(M-1) & y_i^{l-1}(M) & \dots & y_i^{l-1}(M+K-1)
\end{bmatrix}.
\end{aligned}
\tag{12}
$$

Second, we construct a matrix $W_{ik}^l \in R^{M\times K}$ whose rows are repeated copies of $w_{ik} \in R^K$

$$
W_{ik}^l = \begin{bmatrix}
w_{ik}^l(0) & w_{ik}^l(1) & \dots & w_{ik}^l(K-1)\\
\vdots & \vdots & \dots & \vdots\\
w_{ik}^l(0) & w_{ik}^l(1) & \dots & w_{ik}^l(K-1)\\
\vdots & \vdots & \dots & \vdots\\
w_{ik}^l(0) & w_{ik}^l(1) & \dots & w_{ik}^l(K-1)
\end{bmatrix}. \tag{13}
$$

We now consider the Hadamard product of these two matrices

$$
\begin{aligned}
&Y_i^{l-1} \otimes W_{ik}^l\\
&= \begin{bmatrix}
y_i^{l-1}(0)w_{ik}^l(0) & \dots & y_i^{l-1}(K-1)w_{ik}^l(K-1)\\
\vdots & \dots & \vdots\\
y_i^{l-1}(m)w_{ik}^l(0) & \dots & y_i^{l-1}(m+K-1)w_{ik}^l(K-1)\\
\vdots & \dots & \vdots\\
y_i^{l-1}(M-1)w_{ik}^l(0) & \dots & y_i^{l-1}(M+K-1)w_{ik}^l(K-1)
\end{bmatrix}.
\end{aligned}
\tag{14}
$$

Applying the summation operation across rows, we get

$$
\sum\left(Y_i^{l-1} \otimes W_{ik}^l\right)(m) = \sum_{r=0}^{K-1} w_{ik}^l(r)\, y_i^{l-1}(m+r) \tag{15}
$$

which is equivalent to (3). We also note that

$$
\sum Y_i^{l-1} \otimes W_{ik}^l = Y_i^{l-1} w_{ik}^l. \tag{16}
$$

Therefore,

$$
x_{ik}^l(m) = \left(Y_i^{l-1} w_{ik}^l\right)(m) \tag{17}
$$

$$
x_{ik}^l = Y_i^{l-1} w_{ik}^l. \tag{18}
$$

Hence, the 1-D convolution operation can be represented in terms of a single matrix-vector product. This operation lies at the heart of conventional explicit general matrix multiplications (GEMM)-based convolution implementations and enables efficient usage of parallel computational resources, such as GPU cores.

### C. Forward Propagation Through a 1-D Self-ONN Neuron

Equation (11) shows how the Self-ONN formulation of (10) can be represented as a summation of $Q$ individual convolutional operations. Moreover, from (12), a convolutional operation can be represented as a matrix-vector product. We now use these two formulations to represent the transformation of (11) as a single convolution operation, and consequently, a single matrix-vector product, instead of Q-separate ones.

We start by introducing $Y_i^{l-1^{(Q)}} \in \mathbb{R}^{M\times KQ}$ such that

$$
Y_i^{l-1^{(Q)}} = \left[\, Y_i^{l-1} \quad \left(Y_i^{l-1}\right)^{\circ 2} \quad \dots \quad \left(Y_i^{l-1}\right)^{\circ Q} \,\right] \tag{19}
$$

where $\circ n$ is the Hadamard exponentiation operator. The $m$th row of $Y_i^{l-1^{(Q)}}$ can be expressed as

$$
Y_i^{l-1^{(Q)}}(m) = \begin{bmatrix}
y_i^{l-1}(m)\\
\vdots\\
y_i^{l-1}(m+K-1)\\
\vdots\\
y_i^{l-1}(m)^2\\
\vdots\\
y_i^{l-1}(m+K-1)^2\\
\vdots\\
y_i^{l-1}(m)^Q\\
\vdots\\
y_i^{l-1}(m+K-1)^Q
\end{bmatrix}^T. \tag{20}
$$

Moreover, we construct $W_{ik}^{l(Q)} \in R^{M\times KQ}$ by first vectorizing $w_{ik}^{l(Q)} \in R^{K\times Q}$ to $w_{ik}^{\vec{l(Q)}} \in R^{KQ}$ and then concatenating $m$ copies of $w_{ik}^{\vec{l(Q)}}$ along the row dimension, as expressed in the following:

$$w_{ik}^{\vec{l(Q)}} = \begin{bmatrix} w_{ik}^{l(Q)}(0,0) \\ \vdots \\ w_{ik}^{l(Q)}(K-1,0) \\ w_{ik}^{l(Q)}(0,1) \\ \vdots \\ w_{ik}^{l(Q)}(K-1,1) \\ \vdots \\ w_{ik}^{l(Q)}(0,Q-1) \\ \vdots \\ w_{ik}^{l(Q)}(K-1,Q-1) \end{bmatrix}^T \tag{21}$$

$$W_{ik}^{l(Q)}(m) = w_{ik}^{\vec{l(Q)}}. \tag{22}$$

Taking the Hadamard product of $Y_i^{l-1^{(Q)}}$ and $W_{ik}^{l(Q)}$, we get

$$\begin{aligned}&\left(Y_i^{l-1^{(Q)}} \otimes W_{ik}^{l(Q)}\right)(m) \\ &= \begin{bmatrix} y_i^{l-1}(m)w_{ik}^{l(Q)}(0,0) \\ \vdots \\ y_i^{l-1}(m+K-1)w_{ik}^{l(Q)}(K-1,0) \\ \vdots \\ y_i^{l-1}(m)^2 w_{ik}^{l(Q)}(0,1) \\ \vdots \\ y_i^{l-1}(m+K-1)^2 w_{ik}^{l(Q)}(K-1,1) \\ \vdots \\ y_i^{l-1}(m)^Q w_{ik}^{l(Q)}(0,Q-1) \\ \vdots \\ y_i^{l-1}(m+K-1)^Q w_{ik}^{l(Q)}(K-1,Q-1) \end{bmatrix}^T.\end{aligned} \tag{23}$$

Summation of the earlier yields

$$\begin{aligned}&\left(\sum\left(Y_i^{l-1^{(Q)}} \otimes W_{ik}^{l(Q)}\right)\right)(m) \\ &= \sum_{r=0}^{K-1} y_i^{l-1}(m+r)w_{ik}^{l(Q)}(r,0) + \sum_{r=0}^{K-1} y_i^{l-1}(m+r)^2 w_{ik}^{l(Q)}(r,0) \\ &\quad + \cdots + \sum_{r=0}^{K-1} y_i^{l-1}(m+r)^Q w_{ik}^{l(Q)}(r,Q-1) \\ &= \sum_{q=1}^{Q}\sum_{r=0}^{K-1} y_i^{l-1}(m+r)^q w_{ik}^{l(Q)}(r,q-1).\end{aligned} \tag{24}$$

This is equivalent to (10). Therefore, one can now express

$$\left(\sum\left(Y_i^{l-1^{(Q)}} \otimes W_{ik}^{l(Q)}\right)\right)(m) = \widetilde{x_{ik}^{l}}(m). \tag{25}$$

In addition, using (24), we can write

$$\begin{aligned}&\sum\left(Y_i^{l-1^{(Q)}} \otimes W_{ik}^{l(Q)}\right) \\ &= \begin{bmatrix} \sum_{q=1}^{Q}\sum_{r=0}^{K-1} y_i^{l-1}(r)^q w_{ik}^{l(Q)}(r,q-1) \\ \vdots \\ \sum_{q=1}^{Q}\sum_{r=0}^{K-1} y_i^{l-1}(m+r)^q w_{ik}^{l(Q)}(r,q-1) \\ \vdots \\ \sum_{q=1}^{Q}\sum_{r=0}^{K-1} y_i^{l-1}(M-1+r)^q w_{ik}^{l(Q)}(r,q-1) \end{bmatrix}.\end{aligned} \tag{26}$$

Finally, from (25) and (26), we can simply infer that

$$\widetilde{x_{ik}^{l}} = Y_i^{l-1^{(Q)}}\left(w_{ik}^{\vec{l(Q)}}\right). \tag{27}$$

The formulation of (27) provides a key computational benefit, as the forward propagation through the generative neuron is accomplished using a single-matrix-vector multiplication. Hence, in theory, if the computational cost and memory requirement of constructing matrices $Y_i^{l-1^{(Q)}}$ and $W_{ik}^{l(Q)}$ is considered negligible, the complexity of a convolutional neuron is approximately the same as that of the generative neuron, as both can be accomplished by a single-matrix-vector product. Finally, to complete the forward propagation, using (1), we can express

$$\widetilde{x_{k}^{l}} = b_k^l + \sum_{i=0}^{N_{l-1}} \widetilde{x_{ik}^{l}}. \tag{28}$$

### *D. Backpropagation*

We now proceed to derive the BP formulation for the generative neuron model of 1-D Self-ONN by utilizing the vectorized notation introduced in Section II-C. To backpropagate the error through the generative neuron, given the derivative of the loss with respect to the neuron's output, $dL/d\widetilde{x_{ik}^{l}}$, we aim to define $dL/dy_i^{l-1}$, $dL/dw_{ik}^{l(Q)}$, and $dL/db_k^l$.

We start by taking the derivative of (27) w.r.t. $\overline{Y_i^{l-1^{(Q)}}}$ as follows:

$$\frac{d\widetilde{x_{ik}^{l}}(m)}{dY_i^{l-1^{(Q)}}(\overline{m})} = \begin{cases} w_{ik}^{\vec{l(Q)}}, & m = \overline{m} \\ \vec{0}, & \text{otherwise.} \end{cases} \tag{29}$$

Using (29), we can now apply the chain rule to get

$$\frac{dL}{dY_i^{l-1^{(Q)}}} = \frac{dL}{d\widetilde{x_{ik}^{l}}}\frac{d\widetilde{x_{ik}^{l}}}{dY_i^{l-1^{(Q)}}}. \tag{30}$$

Given $(dL/dY_i^{l-1^{(Q)}})$, we aim to find the derivative of the loss w.r.t. to the previous layer's output

$$\frac{dL}{dy_i^{l-1}} = \frac{dL}{dY_i^{l-1}}\frac{dY_i^{l-1}}{dy_i^{l-1}}. \tag{31}$$

We know from (19) that $Y_i^{l-1^{(Q)}} = [Y_i^{l-1} \quad (Y_i^{l-1})^{\circ 2} \quad \ldots \quad (Y_i^{l-1})^{\circ Q}]$. Taking the derivative of (19) w.r.t. $Y_i^{l-1}$

$$\frac{dY_i^{l-1^{(Q)}}}{dY_i^{l-1}} = \left[\mathbf{1} \quad 2\left(Y_i^{l-1}\right)^{\circ 1} \quad \ldots \quad Q\left(Y_i^{l-1}\right)^{\circ Q-1}\right]. \tag{32}$$

Using this, we can write

$$\frac{dL}{dY_i^{l-1}} = \frac{dL}{dY_i^{l-1^{(Q)}}} \frac{dY_i^{l-1^{(Q)}}}{dY_i^{l-1}}. \tag{33}$$

Finally, we can calculate the derivative of the loss w.r.t. $y_i^{l-1}$ as follows:

$$\begin{aligned} \frac{dL}{dy_i^{l-1}(\overline{m})} &= \sum_{m=0}^{M-1} \frac{dL}{dY_i^{l-1}(m)} \frac{dY_i^{l-1}(m)}{dy_i^{l-1}(\overline{m})} \\ &= \sum_{m=0}^{M-1} \frac{dL}{dY_i^{l-1}(m)} \\ &\quad \times \left[\frac{dy_i^{l-1}(m)}{dy_i^{l-1}(\overline{m})}, \ldots, \frac{dy_i^{l-1}(m+K-1)}{dy_i^{l-1}(\overline{m})}\right]. \end{aligned} \tag{34}$$

From (34) and (12), we can notice that $(dY_i^{l-1}(m)/dy_i^{l-1}(\overline{m}))$ will be equal to 1 only when the condition $m \leq \overline{m} \leq (m + K - 1)$ is met, and 0, otherwise. Moreover, as there are no repeating entries in each row of $Y_i^{l-1}$, only one element of $(dY_i^{l-1}(m)/dy_i^{l-1}(\overline{m}))$ can be nonzero and the location of this nonzero element is given by $\mathrm{mod}(\overline{m}, K)$. Based on these two observations, we can infer the following:

$$\begin{aligned} &\frac{dL}{dy_i^{l-1}(\overline{m})} \\ &= \sum_{m=0}^{M-1} \begin{cases} \dfrac{dL}{dY_i^{l-1}}(m, \mathrm{mod}(\overline{m}, K)), & m \leq \overline{m} \leq (m+K-1) \\ 0, & \text{otherwise.} \end{cases} \end{aligned} \tag{35}$$

The only other partial derivative needed for completing the BP is the of the loss w.r.t. the weights of the neuron $w_{ik}^{\overrightarrow{l(Q)}}$. Again, by the chain rule, we can write

$$\frac{dL}{dw_{ik}^{\overrightarrow{l(Q)}}}(\vec{r}) = \frac{dL}{d\widetilde{x_{ik}^l}} \frac{d\widetilde{x_{ik}^l}}{dw_{ik}^{\overrightarrow{l(Q)}}}. \tag{36}$$

where $(d\widetilde{x_{ik}^l}/dw_{ik}^{\overrightarrow{l(Q)}})$ can be calculated by taking the derivative of (27) w.r.t. $w_{ik}^{\overrightarrow{l(Q)}}$ as follows:

$$\frac{d\widetilde{x_{ik}^l}}{dw_{ik}^{\overrightarrow{l(Q)}}} = Y_i^{l-1^{(Q)}}. \tag{37}$$

For the bias, we can use (28) to write

$$\frac{d\mathrm{L}}{db_k^l} = \frac{dL}{d\widetilde{x_k^l}} \frac{d\widetilde{x_k^l}}{db_k^l} = \sum_{m=0}^{M-1} \frac{dL}{d\widetilde{x_k^l}}(m). \tag{38}$$

Finally, assuming a stochastic gradient descent (SGD)-based optimization, the weights and biases can be updated as follows:

$$w_{ik}^{\overrightarrow{l(Q)}}(t+1) = w_{ik}^{\overrightarrow{l(Q)}}(t) - \epsilon(t) \frac{dL}{dw_{ik}^{\overrightarrow{l(Q)}}} \tag{39}$$

$$b_k^l(t+1) = b_k^l(t) - \epsilon(t) \frac{dL}{db_k^l} \tag{40}$$

where $\epsilon(t)$ is the learning factor at iteration $t$.

TABLE I
PATIENT INFORMATION ON THE ECG DATA FROM CPSC-DB

| Patient | AF? | Length(h) | # Tot. Beats | # V Beats | # S Beats |
|---|---|---|---|---|---|
| 1 | No | 25.89 | 109731 | 0 | 24 |
| 2 | Yes | 22.83 | 108297 | 4554 | 0 |
| 3 | Yes | 24.70 | 138878 | 382 | 0 |
| 4 | No | 24.51 | 101734 | 19024 | 3466 |
| 5 | No | 23.57 | 94635 | 1 | 25 |
| 6 | No | 24.59 | 770806 | 0 | 6 |
| 7 | No | 23.11 | 96814 | 15150 | 3481 |
| 8 | Yes | 25.46 | 125495 | 2793 | 0 |
| 9 | No | 25.84 | 89854 | 2 | 1462 |
| 10 | No | 23.64 | 82851 | 169 | 9071 |

## III. METHODOLOGY

In this study, we are aiming to achieve a superior R-peak detection performance compared with the state-of-the-art method in [34] based on a 12-layer 1-D CNN with 448 neurons while reducing the network complexity and depth significantly. Therefore, as shown in Fig. 4, for a fair comparison with [34], the same R-peak approach is followed. However, the architectural complexity of the network is reduced by using a six-layer 1-D Self-ONN with less than 100 neurons. In order to perform fair comparative evaluations against [34], the same network model of [34], the UNet is used for Self-ONNs. As shown in Fig. 4, the peak detection problem is converted to a regression task, which aims to learn the peak locations by transforming the original (normalized) ECG segment into a sequence of five-sample wide pulses. The center of each pulse corresponds to the R-peak location. In order to evaluate its effect, the order of the Taylor polynomials, $Q$, is varied in {3,5,7} over three networks. Each 20-s (8000 samples) ECG segment is linearly normalized in the range, $[-1, 1]$, and then used as input for the 1-D Self-ONNs. The same optimizer (Adam) is used to train the network with the same number of epochs and hyperparameters. The details of the experimental setup and network parameters will be presented in Section IV.

## IV. EXPERIMENTAL RESULTS

In this section, we will first introduce the benchmark ECG dataset, CPSC 2020, used in this study, and then present the experimental setup used for testing and evaluation of the proposed R-peak detector using 1-D Self-ONNs. An extensive set of ECG classification experiments and comparative evaluations against the recent methods over the benchmark CPSC-DB Holter dataset will be presented next. Finally, the computational complexity analysis of 1-D Self-ONNs and 1-D CNNs will be reported in detail.

### *A. China Physiological Signal Challenge-2020*

The CPSC-DB dataset consists of ten single-lead ECG recordings which are collected from arrhythmia patients, each of the recordings lasts for about 24 h. Table I presents the patient information in detail [41].

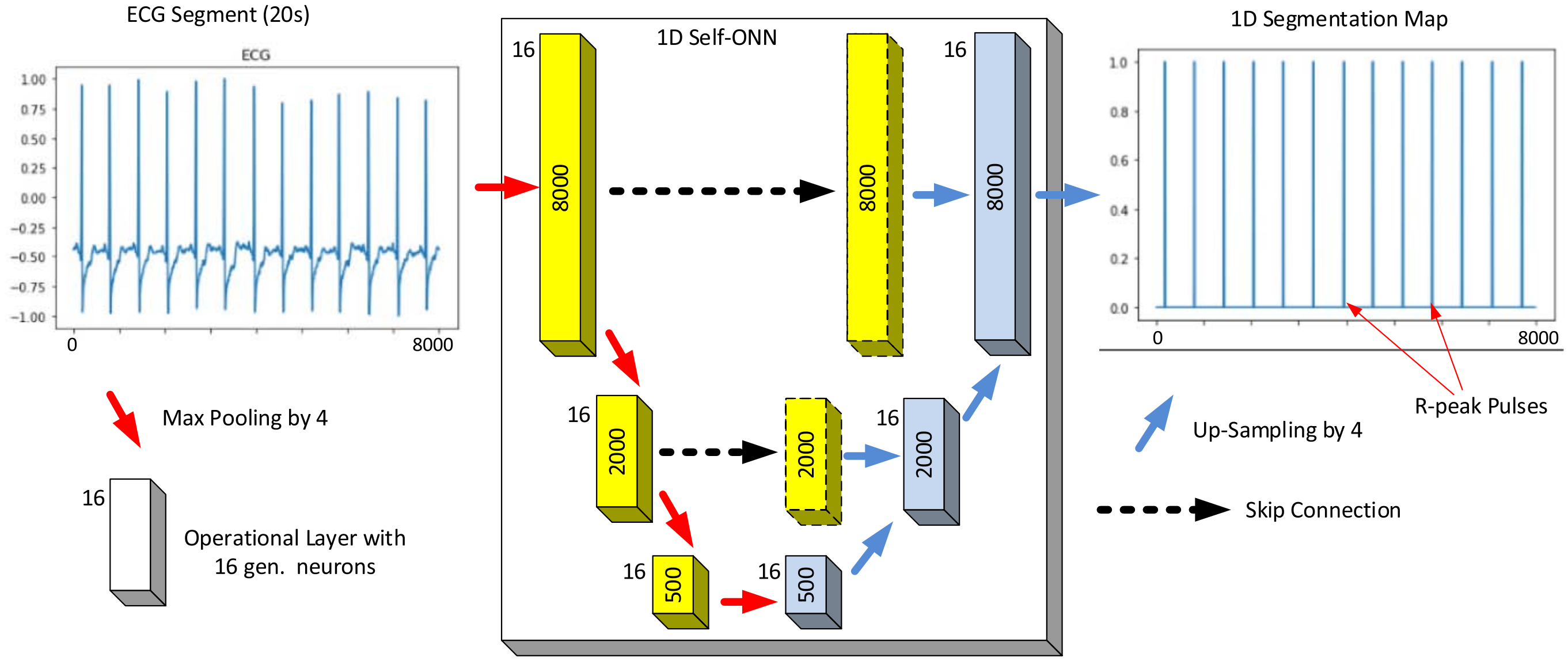

Fig. 4. Proposed approach for R-peak detection.

The other properties of the CPSC-DB (Holter) ECG dataset can be summarized as follows: All ECG data were acquired by a unified wearable ECG device with a sampling frequency of 400 Hz and the total number of beats is 1 026 095. The recordings include irregular heart rhythms and supraventricular premature beats (SPBs or S beats) and premature ventricular contraction (PVC or V) type beats. All recordings are provided in MATLAB format with corresponding S and V beats annotations. R-peak annotations for each ECG cycle were annotated by a team of biomedical researchers. To show the robustness of the R-peak detector against noise and other artifacts, CPSC-DB presents a real-world Holter dataset containing numerous ECG containments and artifacts.

### B. Experimental Setup

For all experiments, as in [34], shallow training is employed where the number of BP epochs is limited to 50. We set the learning rate, $\epsilon(0)$, as $10^{-3}$ and an Adam optimizer is used for minimizing the binary cross-entropy loss (BCE). We performed three individual BP runs for training over each patient's data, and for comparative evaluations, we report the best detection performance.

The comparative evaluations of the proposed R-peak detector based on 1-D Self-ONNs are carried out against the following two detectors with 1-D CNNs: 1) the deep 1-D CNN from [34], and 2) 1-D CNN with the same network configuration as 1-D Self-ONN. Moreover, comparative evaluations are also performed against five earlier *state-of-the-art* classifiers from the literature [23], [56]–[59]. R-peak detection experiments are performed over the CSPC dataset. Each classifier is trained over nine patients and tested over the (unseen) patient. Therefore, tenfold cross validation is performed for comparative evaluations over all ten patients with overall 1 026 095 beats. Over the cumulated TP, FP, and FN counters, and the standard performance metrics, *Precision* or *Positive Predictivity* ( *Ppr*), *Recall* or *Sensitivity* (*Sen*), and $F1$-*score*, which is the harmonic mean of the model's *Precision* and *Recall* are computed. The true-negative (TN) is omitted in peak detection (TN = 0). The expressions of these performance metrics using the hit/miss counters, e.g., TP, TN, FP, and FN, are as follows:

$$\text{Ppr} = \frac{\text{TP}}{\text{TP}+\text{FP}}, \quad \text{Sen} = \frac{\text{TP}}{\text{TP}+\text{FN}}, \quad F1 = \frac{2\text{PprSen}}{\text{Ppr}+\text{Sen}}. \qquad (41)$$

### C. Peak Detection Performance Evaluation

R-peak detection results over the entire CSPC dataset are presented in Table II. The (performance) loss (%) in FN (or FP or other "miss" metrics) of a method $X$ compared with the method with the minimum FN can be defined as follows:

$$\text{FN LOSS}(X)\% = 100\frac{\text{FN(min)}}{\text{FN}(X)}. \qquad (42)$$

It is clear that 1-D Self-ONNs achieved a significant performance gap when compared with the 1-D CNNs with the same number of learning units (neurons) and depth. Overall, Self-ONNs with $Q = 3$ and $Q = 5$ reduced the FPs and FNs on peak detection by more than 43% and 64%, respectively. Even when compared with the state-of-the-art 1-D CNN configuration in [34] with twice the depth and more than four times the number of neurons, the detection errors, FPs and FNs, were reduced by more than 7% and 37%, respectively. Such a substantial reduction especially on FNs over both 1-D CNN configurations shows that 1-D Self-ONNs can indeed accomplish a superior learning capability to detect the actual peaks. The performance loss is significantly higher for the earlier methods [23], [56]–[59]. Finally, the results indicate that the best performances are obtained by 1-D Self-ONNs with either $Q = 3$ or $Q = 5$. We have found that the former setting is a better choice because of the higher F1 score, significantly lower FPs, and lower computational cost.

TABLE II

OVERALL PEAK DETECTION PERFORMANCE OF THE CLASSIFIERS. 1-D CNN (*) AND 1-D SELF-ONN HAVE THE SAME CONFIGURATION. THE BEST RESULTS ARE PRESENTED IN **Bold**

| **Network** | ***Q*** | **TP** | **FN** | **FN LOSS (%)** | **FP** | **FP LOSS (%)** | ***Sen*** | ***Ppr*** | **F1** |
|---|---|---|---|---|---|---|---|---|---|
| LSTM [56] | **-** | 1,007,823 | 18,272 | 89.02 | 23,835 | 45.88 | 98.20 | 97.76 | 97.88 |
| P and T [23] | **-** | 998,413 | 27,682 | 92.75 | 28,940 | 55.43 | 97.31 | 97.18 | 97.23 |
| Hamilton [57] | **-** | 993,920 | 32,175 | 93.76 | 62,733 | 79.44 | 96.82 | 93.73 | 95.14 |
| Two M. Avg. [58] | **-** | 992,305 | 33,790 | 94.06 | 46,349 | 72.17 | 96.66 | 95.34 | 95.97 |
| SWT [59] | **-** | 975,222 | 50,873 | 96.05 | 27,936 | 53.83 | 95.06 | 97.20 | 96.09 |
| **1D CNN [34]** | **1** | 1,022,874 | 3221 | 37.69 | 13931 | 7.41 | 99.70 | 98.63 | 99.16 |
| **1D CNN (*)** | **1** | 1,020,544 | 5651 | 64.48 | 22820 | 43.48 | 99.43 | 97.83 | 98.57 |
| **1D Self-ONN** | **3** | 1,023,997 | 2098 | 4.34 | **12899** | **0.00** | 99.80 | **98.77** | **99.28** |
| | **5** | **1,024,088** | **2007** | **0.00** | 14263 | 9.56 | **99.81** | 98.63 | 99.21 |
| | **7** | 1,023,907 | 2188 | 8.27 | 16481 | 21.73 | 99.79 | 98.42 | 99.10 |

TABLE III

PEAK DETECTION PERFORMANCE OF THE CLASSIFIERS OVER THE ARRHYTHMIA (S AND V) BEATS. 1-D CNN (*) AND 1-D SELF-ONN HAVE THE SAME CONFIGURATION. THE BEST RESULTS ARE PRESENTED IN **Bold**

| | | **S beats** | | | **V beats** | | |
|---|---|---|---|---|---|---|---|
| **Network** | **Q** | **Detected** | **Miss** | **LOSS (%)** | **Detected** | **Miss** | **LOSS (%)** |
| LSTM [56] | **-** | 17,330 | 205 | 90.73 | 29,247 | 12,828 | 98.21 |
| P and T [23] | **-** | 17,230 | 305 | 93.77 | 31,369 | 10,706 | 97.85 |
| Hamilton [57] | **-** | 17,163 | 372 | 94.89 | 31,906 | 10,169 | 97.74 |
| Two M. Avg. [58] | **-** | 17,231 | 304 | 93.75 | 28,972 | 13,103 | 98.44 |
| SWT [59] | **-** | 17,018 | 517 | 96.32 | 27,283 | 14,792 | |
| **1D CNN [34]** | **1** | 17,495 | 40 | 52.50 | 41,523 | 552 | 58.33 |
| **1D CNN (*)** | **1** | 17,463 | 72 | 73.61 | 41,687 | 738 | 68.83 |
| **1D Self-ONN** | **3** | 17,504 | 31 | 38.71 | **41,805** | **230** | **0.00** |
| | **5** | **17,516** | **19** | **0.00** | 41,755 | 235 | 2.13 |
| | **7** | 17,507 | 28 | 32.14 | 41,713 | 287 | 19.86 |

In peak detection, detecting the arrhythmia, i.e., supraventricular premature (S) and premature ventricular contraction (V) beats is crucial since R-peak detection is the prior operation to an automated ECG beat classification and arrhythmia detection. Obviously, this aim cannot be fulfilled if the peak detector fails to detect an abnormal beat. For this purpose, the focus is then drawn on the peak detection performance over the arrhythmia (S and V) beats. As shown in Table III, once again 1-D Self-ONNs have shown superiority for detecting both S and V beats over both 1-D CNN configurations. This time even the deep 1-D CNN causes over 52% and 58% more misdetections, respectively, and again, the gap further widens over the 1-D CNN with the same configuration. Finally, the loss of the earlier methods [23], [56]–[59] has peaked above 90%, which shows that those methods are not robust at all for the Holter ECG.

As for visual comparison, over the four ECG segments, Fig. 5 shows R-peak detection results of 1-D Self-ONNs and the two 1-D CNN configurations. Such typical visual results clearly show that both 1-D CNN models yield numerous FPs and FNs especially when the R-peak is corrupted with high noise or some abrupt change occurring in the close vicinity, e.g., abrupt shifts on the baseline or occasional voltage glitches. Especially, the deep CNN model [34] missed both arrhythmic peak locations in Fig. 5(d) which is a substantial error.

### *D. Computational Complexity*

In this section, we provide the formulation for calculating the total number of multiply-accumulate operations (MACs) and the total number of parameters (PARs) of a generative neuron of a 1-D Self-ONN. To calculate the number of trainable parameters, we recall from Section II that, for each kernel connection, the generative neuron has $Q$ times more learnable parameters. Cumulatively, the number of trainable parameters, $n_k^l$, of the $k$th neuron of $l$th layer is given by the following formulation:

$$n_k^l = N_{l-1} * K_k^l * Q_k^l. \tag{43}$$

In (43), $N_{l-1}$ is the number of neurons in layer $l-1$, $K_k^l$ is the kernel size used in the neuron, and $Q_k^l$ is the approximation order selected for this neuron. Finally, to calculate the total number of MAC operations, one can note from (26) that to produce a single element in the output $\widetilde{x_{ik}^l}$, we require $K_k^l * Q_k^l$ MAC operations for each output map $y_i^{l-1}$ of the previous

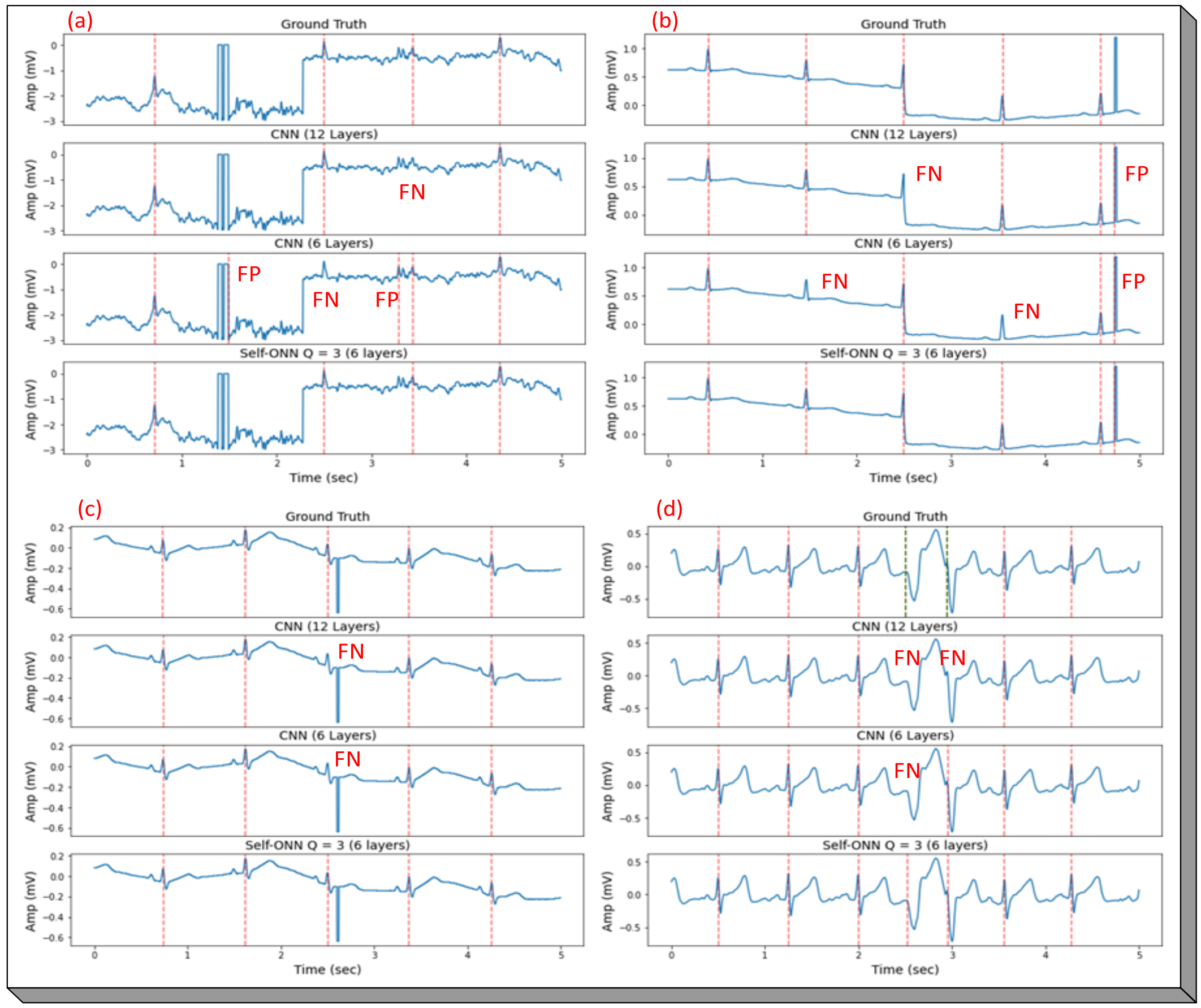

Fig. 5. (a)–(d) Visualization of R-peak detection results over four ECG segments. For each segment, (top to bottom) rows show the ground truth, the output of the 12-layer CNN ([34]), the output of the 6-layer CNN version of [34], and the output of the proposed 1-D Self-ONN. Typical FPs and FNs are visible over the detections of both 1-D CNN models. Ground truth peak locations are shown on the top plot where red and black dashed lines correspond to normal and arrhythmic peak locations, respectively.

TABLE IV

NETWORK MODELS AND THEIR COMPUTATIONAL COMPLEXITIES. THE AVERAGE TIME CORRESPONDS TO THE TIME TO DETECT R-PEAK LOCATIONS OF A 20-s ECG SEGMENT

| Network | *PARs* (*K*) | *MACs* (M) | Avg. Time (ms) | | Train Time (min) |
|---|---|---|---|---|---|
| | | | CPU | GPU | |
| 1D CNN **[34]** | 79,121 | 67.05 | 191 | 110 | 37.0 |
| 1D CNN (*) | 12,801 | 46.36 | 90 | 64 | 11.8 |
| 1D Self-ONN (Q=3) | 38,209 | 138.65 | 106 | 66 | 19.5 |

layer. Generalizing this, we can write the following:

$$\mathrm{MAC}_k^l = N_{l-1} * |\widetilde{x_{ik}^l}| * K_k^l * Q_k^l \tag{44}$$

where $|\cdot|$ is the cardinality operator. For notational convenience, the bias term and the cost of the Hadamard exponentiation are omitted from (44).

We implemented the proposed 1-D Self-ONN using Python and FastONN [61] library, based on PyTorch [62]. All the experiments reported in this article were run on a 2.2-GHz Intel Core i7-8750H with 16 GB of RAM and an NVIDIA GeForce GTX 1060 graphic card. Both training and evaluation of the classifier were processed by CUDA kernels. Along with the average time complexity, using the formulations in (43) and (44), we provide the overall *PARs* and *MACs* for both network models in Table IV. As shown in Table IV, a significant gap occurs between 1-D Self-ONN and deep CNN models in terms of the total number of parameters and average computation time. The Self-ONN network requires the most number of multiply-accumulate operations. However, a higher percentage of these operations are independent, and thus, parallelizable. Therefore, an efficient implementation using the formulation of (27) results in an actual running time that is less than the deep CNN and only marginally (3.1%) higher than the equivalent 1-D CNN, which is negligible considering the crucial gain in the detection performance. The optimized PyTorch implementation of Self-ONNs is publicly shared in [63].

## V. CONCLUSION

In this study, 1-D Self-ONNs are proposed for R-peak detection, especially for poor quality ECG signals, e.g., acquired by Holter monitors or low-power mobile ECG sensors. The primary goal is to achieve the state-of-the-art R-peak detection performance with an elegant computational efficiency for a real-time application over such low-power ECG devices. As a new-generation network model, a Self-ONN is a highly heterogeneous network composed of generative neurons. This yields a crucial advantage of optimizing the nodal operator function of each kernel element, and thus, Self-ONNs can achieve an utmost heterogeneity that maximizes the network diversity and the learning performance. As a result, the traditional weight optimization of conventional CNNs is entirely turned into an operator generation process via optimization. Despite its highly nonlinear kernel elements, each Self-ONN layer can still be implemented by a single 1-D convolution, and this allows a parallelized implementation similar to the one for conventional CNNs.

We performed tenfold comparative evaluations over the benchmark CPSC dataset with more than 1M beats. Against the current *state-of-the-art* method proposed in [34] with a 12-layer CNN, 1-D Self-ONNs significantly reduced both FPs and FNs even though it has *half* the depth and more than four times fewer neurons. Against the 1-D CNN with the equivalent configuration, the performance gap further widens. The most crucial advantage is that 1-D Self-ONNs can reduce more than 52% and 58% of the overall misdetections of the S and V arrhythmia beats, respectively, compared with the deep CNNs. Finally, the 1-D Self-ONN model used in this study presents a superior computational efficiency with respect to the deep 1-D CNNs, and thus, especially for low-power, mobile devices, such as Holter monitors, the proposed approach can conveniently be used as an R-peak detector in real-time. The optimized implementation of the proposed peak detector and the benchmark CPSC dataset is shared in [64].